\pdfoutput=1

\documentclass[aps,pra, amsmath,superscriptaddress,floatfix,twocolumn,longbibliography, 10pt]{revtex4-1}

\usepackage{graphicx}   
\usepackage{siunitx}
\usepackage{bm}
\usepackage{physics}

\begin{document}

\title{Valley-selective optical Stark effect probed by Kerr rotation}
\author{Trevor LaMountain}
\affiliation{Applied Physics Program, Northwestern University, Evanston, Illinois 60208, USA}
\author{Hadallia Bergeron}
\affiliation{Department of Materials Science and Engineering, Northwestern University, Evanston, Illinois 60208, USA}
\author{Itamar Balla}
\affiliation{Department of Materials Science and Engineering, Northwestern University, Evanston, Illinois 60208, USA}
\author{Teodor K. Stanev}
\affiliation{Department of Physics and Astronomy, Northwestern University, Evanston, Illinois 60208, USA}
\author{Mark C. Hersam}
\affiliation{Applied Physics Program, Northwestern University, Evanston, Illinois 60208, USA}
\affiliation{Department of Materials Science and Engineering, Northwestern University, Evanston, Illinois 60208, USA}
\affiliation{Department of Chemistry, Northwestern University, Evanston, Illinois 60208, USA}
\affiliation{Department of Electrical Engineering and Computer Science, Northwestern University, Evanston, Illinois 60208, USA}
\author{Nathaniel P. Stern}\
\email{n-stern@northwestern.edu}
\affiliation{Applied Physics Program, Northwestern University, Evanston, Illinois 60208, USA}
\affiliation{Department of Physics and Astronomy, Northwestern University, Evanston, Illinois 60208, USA}

\date{\today}

\begin{abstract}

The ability to monitor and control distinct states is at the heart of emerging quantum technologies.  The valley pseudospin in transition metal dichalcogenide (TMDC) monolayers is a promising degree of freedom for such control, with the optical Stark effect allowing for valley-selective manipulation of energy levels in WS$_2$ and WSe$_2$ using ultrafast optical pulses.  Despite these advances, understanding of valley-sensitive optical Stark shifts in TMDCs has been limited by reflectance-based detection methods where the signal is small and prone to background effects. More sensitive polarization-based spectroscopy is required to better probe ultrafast Stark shifts for all-optical manipulation of valley energy levels.  Here, we show time-resolved Kerr rotation to be a more sensitive probe of the valley-selective optical Stark effect in monolayer TMDCs. Compared to the established time-resolved reflectance methods, Kerr rotation is less sensitive to background effects. Kerr rotation provides a five-fold improvement in the signal-to-noise ratio of the Stark effect optical signal and a more precise estimate of the energy shift. This increased sensitivity allows for observation of an optical Stark shift in monolayer MoS$_2$ that exhibits both valley- and energy-selectivity, demonstrating the promise of this method for investigating this effect in other layered materials and heterostructures.

\end{abstract}

\maketitle

\section{Introduction}
Manipulating electronic systems with light provides a controllable, high-speed, and non-destructive mechanism for coherent measurement or control of quantum states~\cite{unold2004optical,awschalom2002optical,gupta2001ultrafast,press2008complete,berezovsky2008picosecond,mikkelsen2009ultrafast}.  First observed in atomic and molecular systems~\cite{autler1955stark,bonch1968current,liao1975direct,verkerk1985doppler}, off-resonant light can cause energy levels to shift with negligible optical absorption, establishing the basic mechanism underlying neutral atom dipole traps~\cite{chu1986experimental,raab1987trapping}. This light-matter interaction, known as the optical Stark effect, can also be used in solid state systems for ultrafast manipulation of semiconductor exciton levels in III-V quantum wells and quantum dots~\cite{mysyrowicz1986dressed,von1986optical,knox1989femtosecond,chemla1989excitonic}. By harnessing the optical selection rules of materials, the Stark effect can be selectively applied to spin-polarized excitons~\cite{joffre1989laser,sieh1999coulomb,combescot1990optical}, thereby splitting degenerate spin states without the use of external magnetic fields. This capability for all-optical state shifting can enable both classical applications such as ultrafast switches ~\cite{gansen2002femtosecond} and modulators~\cite{jin1990ultrafast} as well as quantum coherent control of spin ~\cite{unold2004optical,awschalom2002optical,gupta2001ultrafast,press2008complete,berezovsky2008picosecond,mikkelsen2009ultrafast}. The polarization-sensitive optical Stark effect can be generalized for coherent control of more exotic systems, evidenced most recently by the valley pseudospin in monolayer transition metal dichalcogenides (TMDCs)~\cite{ye2016optical}.

In monolayer TMDC crystals, inversion asymmetry and spin-orbit coupling create two regions of momentum space, or valleys, that are degenerate in energy but possess different Berry curvature and coupling to circularly polarized light. The resulting valley pseudospin becomes an intrinsic label   for the distinct electronic~\cite{xiao2012coupled}, excitonic~\cite{Xu2014}, and recently, polaritonic~\cite{Chen2017} excitations in monolayer TMDCs that can potentially be used to carry information in analogy to spin~\cite{schaibley2016valleytronics}. Using circularly polarized light, the optical Stark effect can selectively break the valley degeneracy~\cite{kim2014ultrafast,sie2015}. Exploiting  the polarization-selective optical Stark effect for coherent control of the valley degree of freedom~\cite{ye2016optical} demonstrates a key requirement for information processing in valleytronics, but remains a relatively unexplored phenomenon observed in only a handful of monolayer TMDCs.

The ability to precisely monitor the evolution of valley peudospin is essential for development of coherent valleytronics. To date, the optical Stark shift ($\Delta E$) of one valley has typically been monitored with absorption-sensitive probes~\cite{kim2014ultrafast,sie2015,ye2016optical}, but these approaches are hindered by a variety of absorptive background effects~\cite{von1986optical,mysyrowicz1986dressed,kim2014ultrafast,sie2015,ye2016optical} that can pollute the signal and limit sensitivity. By using optical probes directly sensitive to polarization, polarization-independent background signals can be suppressed, enabling more precise measurements of the valley splitting induced by the optical Stark effect in a broader set of materials and heterostructures.

\begin{figure*}[tb]
\centering
\includegraphics{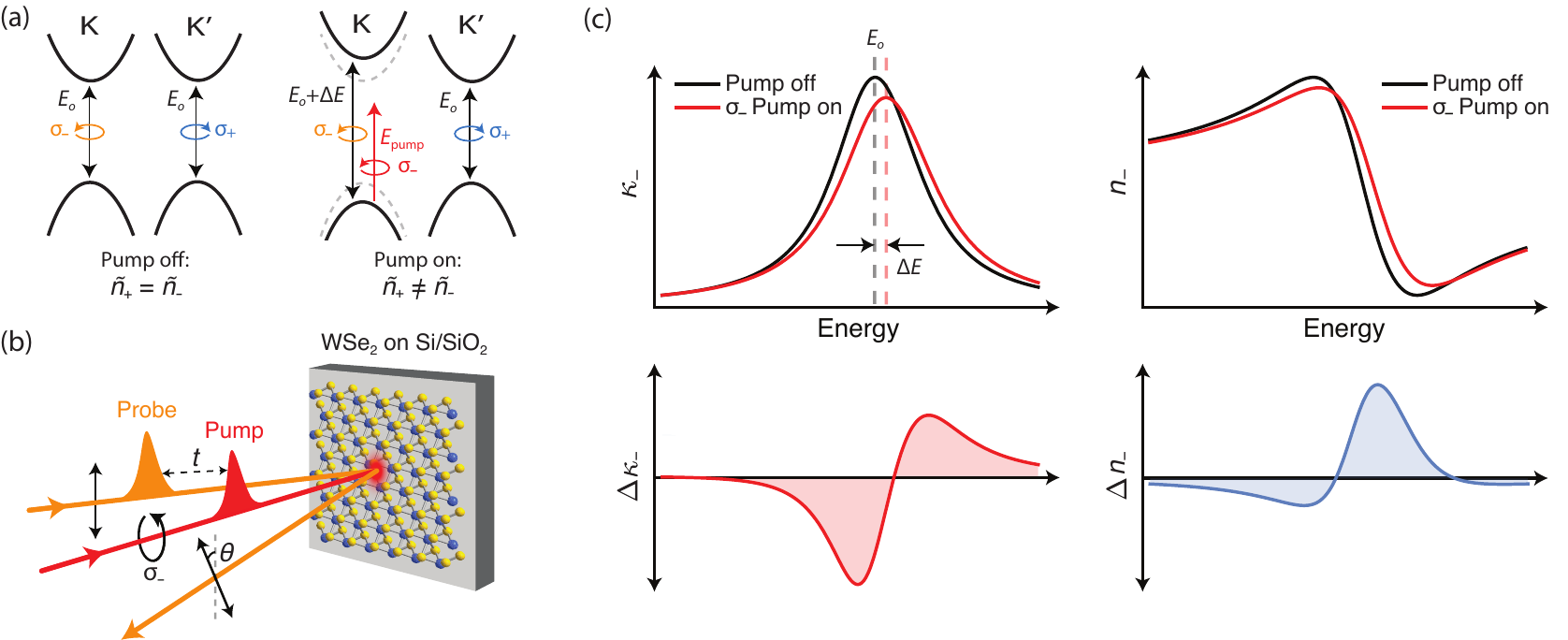}
\caption{
Kerr rotation induced by the valley-selective optical Stark effect.  (a) The valley Stark shift in the monolayer TMDC band structure. Due to optical selection rules, a $\sigma_-$ below-bandgap pump only couples to the K valley, resulting in a valley-dependent optical Stark shift and corresponding change in the complex refractive index $\tilde{n}_-$. (b) Schematic of TR-KR on monolayer WSe$_2$. A detuned circularly-polarized pump induces a rotation of the linear polarization angle $\theta$ in the reflected probe when the beams are spatially and temporally overlapped ($t = 0$). (c) Illustration of the imaginary ($\kappa_-$) and real ($n_-$) parts of the refractive index in WSe$_2$ near the exciton resonance. The sub-bandgap pump causes a blue-shift $\Delta E$ of the exciton in the K valley, which changes the $\kappa_-(\omega)$ and $n_-(\omega)$ spectra. The bottom panels show the corresponding spectra of $\Delta \kappa_-(\omega)$ and $\Delta n_-(\omega)$, which are defined as the difference between the spectra with the $\sigma_ -$ pump on and off. The $\sigma_-$ pump does not couple to the K$^\prime$ valley, so $\tilde{n}_+$ is unaffected.}
\label{fig:OpticalStarkCartoon}
\end{figure*}

Taking advantage of polarization selection rules, time-resolved Kerr and Faraday rotation are powerful methods for probing coherent rotation of spin states caused by the optical Stark effect in semiconductors ~\cite{gupta2001ultrafast,gupta2001spin,mikkelsen2009ultrafast,zhang2010tailoring}.  Similar selection rules at the K and K$^\prime$ valleys of single-layer TMDCs dictate that the band-edge optical transitions in each valley uniquely couple to left-handed ($\sigma_-$) or right-handed ($\sigma_+$) circularly polarized light, respectively~\cite{xiao2012coupled} (Fig.~\ref{fig:OpticalStarkCartoon}a). Asymmetry in the response of monolayer TMDCs to $\sigma_-$ and $\sigma_+$ light, encoded in the complex index of refraction ($\tilde{n}_{\pm} = n_{\pm} + i \kappa_{\pm}$), can be detected by the Kerr rotation angle $\theta$ of reflected linearly polarized light (Fig.~\ref{fig:OpticalStarkCartoon}b). Using an above-bandgap pump, time-resolved Kerr rotation (TR-KR) can be used to measure valley lifetimes of carriers in TMDCs~\cite{zhu2014exciton,plechinger2014time,song2016long,huang2017temporal,dey2017gate}. Since a valley-selective energy shift also generates an asymmetry between $\tilde{n}_{\pm}$~\cite{kim2014ultrafast,sie2015,ye2016optical}~(Fig.~\ref{fig:OpticalStarkCartoon}c), TR-KR should be similarly sensitive to the valley-selective optical Stark effect caused by an off-resonant below-bandgap pump.

In this paper, we demonstrate the use of time-resolved Kerr rotation to measure the valley-selective optical Stark effect in TMDC monolayers. The established time-resolved reflectance (TR-R)~\cite{kim2014ultrafast,sie2015,ye2016optical} measurement is performed on the same sample under the same pump and probe conditions to provide quantitative comparison of the two methods. $\Delta E$ extracted from TR-R and TR-KR for WSe$_2$ agree, confirming the accuracy of the analysis. TR-KR is free from the absorptive background signals present in TR-R, allowing for more direct and precise measurement of the valley-selective Stark shift.
With a nominal rotation angle sensitivity of 3 \si{\micro\radian}, we use TR-KR to effectively measure a Stark shift of 4 \si{\micro eV} -- the smallest reported shift in current literature.   We exploit the increased sensitivity and background signal suppression of TR-KR to observe the valley- and energy-selective optical Stark effect in MoS$_2$.

\section{Pump-probe measurements of the optical Stark effect}
Both time-resolved reflectance and Kerr rotation measurements of the valley-selective optical Stark effect use ultrafast pulses to induce an excitonic energy shift and to detect the pump-induced changes in $\tilde{n}_\pm$ as a function of pump-probe delay time $t$.  TR-R measures the change in reflectance of one particular valley using a circularly-polarized probe. In contrast, TR-KR measures the induced phase difference between circular polarization components of a linearly polarized probe.  TR-R and TR-KR measurements predominantly probe different parts of the refractive index: for transparent substrates, TR-R is primarily sensitive to changes in the imaginary part ($\Delta \kappa$)~\cite{von1986optical,mysyrowicz1986dressed,kim2014ultrafast,sie2015,ye2016optical}, while TR-KR is primarily sensitive to changes in the real part ($\Delta n$)~\cite{kuch2015magnetic}. While the exact relationship between the TR-R/TR-KR spectra and the complex refractive index is more complicated with reflective substrates due to thin film effects, the complementary nature of the two measurements persists (see Supplementary Material (SM)). Regardless of substrate, Kerr rotation probes a part of the $\Delta \tilde{n}$ spectrum induced by the valley-selective optical Stark effect that was previously unmeasured using absorptive methods.

The magnitude of the Stark shift in the case of a continuous-wave pumping field is given by
\begin{equation}
\Delta E_{\rm cw} = \frac{|\mathcal{M}|^2 \mathcal{E}^2 } {2 \Delta_{\rm pump}}
\label{eq:StarkContinuous}
\end{equation}
where $\mathcal{M}$ is the transition dipole moment of the exciton transition, $ \mathcal{E}$ is the amplitude of the pump electric field, and $\Delta_{\text{pump}} =  E_0 - E_{\text{pump}}$ is the red-detuning of the pump energy $E_{\rm pump}$ from the exciton resonance $E_0$~\cite{sie2015}. In ultrafast experiments, the observed Stark shift depends on the pump-probe delay $t$ and has a maximum value when the pump and probe pulses overlap in time ($t=0$). The comparable duration of the pump and probe pulses means that the probe samples a range of pump field amplitudes, so the observed maximum Stark shift is actually a convolution of the temporal profile of the Stark shift and the probe pulse. Moreover, the competing timescales of pump pulse width and exciton formation can significantly reduce the magnitude of the observed Stark shift~\cite{ye2016optical}. Hence, Eq.~\eqref{eq:StarkContinuous} is not directly applicable to $\Delta E$ measured by pump-probe techniques.  Instead, the magnitude of the observed Stark shift is characterized by
\begin{equation}
\Delta E = C \frac{I_{\rm peak}}{\Delta_{\rm pump}}
\label{eq:StarkPulsed}
\end{equation}
where $I_{\rm peak} \propto \mathcal{E}_{\rm peak}^2$ is the peak intensity of the pump pulse, and $C$ is  a phenomenological proportionality constant. $C$ accounts for deviations from Eq.~\eqref{eq:StarkContinuous} due to exciton formation time and pulse width, and also absorbs $\vert \mathcal{M} \vert^2$.

\section{Increased sensitivity to Stark shift using Kerr Rotation}
The optical Stark effect has previously been observed in WSe$_2$ and WS$_2$ using TR-R, but an absorptive background signal is present in all reported measurements~\cite{kim2014ultrafast,sie2015,ye2016optical}. Here, we measure the optical Stark effect in WSe$_2$ on a Si/SiO$_2$ substrate using both TR-R and TR-KR to demonstrate not only the suppression of these background signals when using TR-KR, but also increased sensitivity to the Stark shift $\Delta E$. Conditions for TR-R and TR-KR experiments are identical for accurate comparisons.  All measurements are performed under vacuum at 20~K.  Pump and probe pulses have an effective full-width at half-maximum (FWHM) of $\Delta t \sim 375$ fs estimated from the observed temporal response (see SM). The pump has center energy $E_{\rm pump} = 1.49$~eV. A peak pump intensity $I_{\rm peak} = \rm fluence/ \Delta t$ = 320~\si{MW/cm^2} is used for all measurements unless otherwise noted. Probe pulses have peak intensity $I_{\rm probe} \sim 4$~\si{MW/cm^2}. At 20~K, the lowest-energy exciton in WSe$_2$ has energy $E_0$ = 1.73~eV, and $\Delta_{\rm pump} = 0.24$~eV is small enough that the Bloch-Seigert shift can be ignored~\cite{cohen1973quantum,sie2017}. The probe is tuned over a range of energies near $E_0$ from 1.68 -- 1.8~eV to obtain the TR-R and TR-KR spectra, with uncertainty estimated by repeating the measurements three times at each probe energy.

\subsection{Time-resolved reflectance}

\begin{figure}[tb]
\centering
\includegraphics{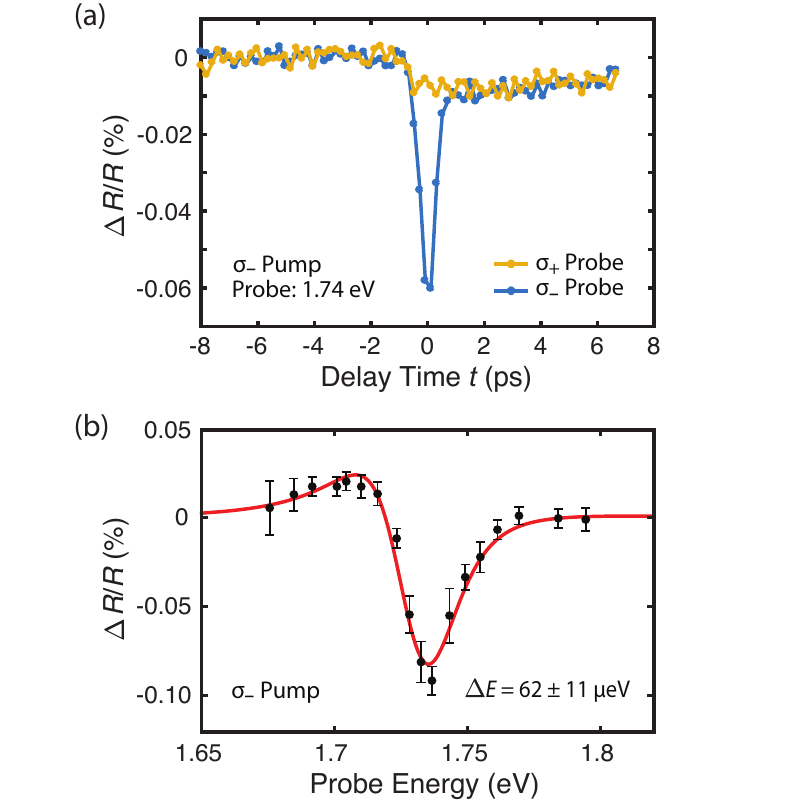}
\caption{Optical Stark effect in WSe$_2$ probed by TR-R.  (a) Differential reflectance as a function of probe delay $t$ for monolayer WSe$_2$ at 20~K using both $\sigma_-$ and $\sigma_+$ polarized probes. The background signal at $t > 0$ is present for both probe polarizations, while the optical Stark effect signal near $t = 0$ only occurs when the pump and probe polarizations have the same helicity. (b) Spectral dependence of the TR-R optical Stark signal after background subtraction. The fit (red) allows extraction of the Stark shift $\Delta E$.}
\label{fig:WSe2Reflectivity}
\end{figure}

We first measure the transient differential reflectance $\Delta R/R$ to quantify the influence of background signals and to establish the precision of TR-R using our experimental setup. Fig.~\ref{fig:WSe2Reflectivity}a shows the probe polarization dependence of TR-R at $E_0 = 1.74$~eV for a $\sigma_-$ pump. Since the exciton energy only blue-shifts during the pump pulse, the signal of interest only occurs when the pump and probe pulses overlap. This manifests as a peak in $\Delta R/R$ of width $\sim500$~fs near $t = 0$ only when probe and pump are co-polarized ($\sigma_-$), indicative of the valley-selective optical Stark effect. There is also a signal that is independent of probe polarization that persists beyond $t\approx 0$. As previously reported~\cite{knox1989femtosecond,sie2015,sim2016selectively}, this background can be explained by two-photon excitation of real excitons by the pump. The two-photon excitation energy (2.99~eV) is significantly larger than the exciton bandgap (1.73~eV) at the K points. Since the valley selection rules are only valid near the K points of the band structure~\cite{yao2008valley,xiao2012coupled}, the hot carriers generated by two-photon absorption at 2.99~eV are not valley specific. Their relaxation creates equal exciton populations in both valleys, resulting in a TR-R background signal that is independent of the probe polarization.

The TR-R spectrum shown in Fig.~\ref{fig:WSe2Reflectivity}b is the difference between the $t=0$ signals for the $\sigma_-$ and $\sigma_+$ probe polarizations as a function of probe energy. This represents the circular dichroism spectrum induced by the Stark shift. The TR-R spectrum does not exhibit the ``Lorentzian-derivative" lineshape reported for WSe$_2$ on sapphire~\cite{kim2014ultrafast} due to the reflective substrate. $\Delta E$ is extracted from the TR-R spectrum following the analysis of Ref.~\cite{ye2016optical} by modeling the exciton resonance as a single Lorentz oscillator. From the TR-R analysis, we find $\Delta E = 62 \pm 11$~\si{\micro eV}, where the range represents the 95\% confidence interval of the estimate (see SM).

\subsection{Time-resolved Kerr rotation}
\begin{figure}[tb]
\centering
\includegraphics{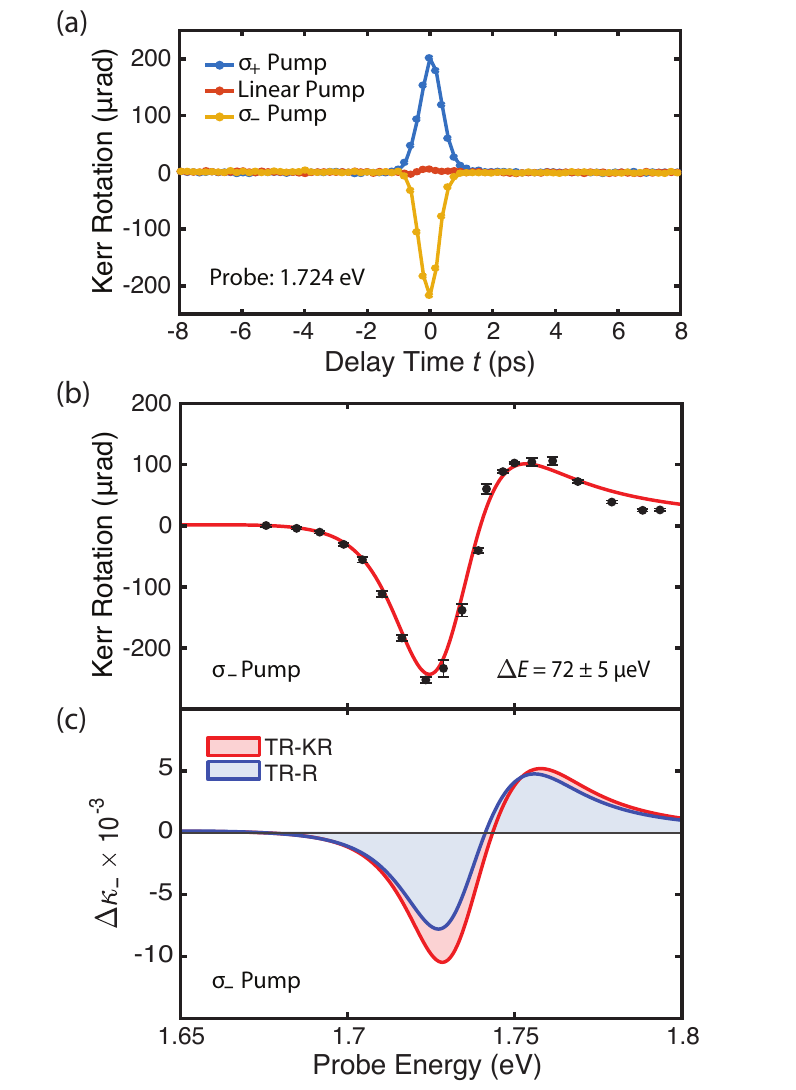}
\caption{Optical Stark effect in WSe$_2$ probed by TR-KR. (a) Kerr rotation angle as a function of delay time $t$ in monolayer WSe$_2$ at 20~K for different pump polarizations. The instantaneous signal near $t=0$ for circularly polarized pump light is caused by the optical Stark effect. (b) Spectral dependence of TR-KR  for $\sigma$- pump. (c) $\Delta \kappa_-$ spectra extracted from fits to the TR-KR and TR-R spectra for the same pump conditions. }
\label{fig:WSe2Kerr}
\end{figure}

We use the same pump-probe apparatus with modified polarization and detection optics to measure the valley-selective Stark shift using Kerr rotation. Fig.~\ref{fig:WSe2Kerr}a shows the pump polarization dependence of the TR-KR signal $\theta(t)$.  Once again, there is a signal near $t=0$, but now the background signal that was present in TR-R is suppressed below noise levels for all pump powers available to our system. Upon switching pump polarization from $\sigma_-$ to $\sigma_+$, the Kerr rotation angle has comparable magnitude but inverted sign. There is no appreciable signal for a linearly polarized pump. This is in contrast to TR-R with a linearly polarized pump, which shows a signal in both valleys with half the magnitude as that observed with the circularly polarized pump~\cite{sie2015}. This distinction is because Kerr rotation probes the \textit{difference} in the refractive index of the WSe$_2$ for $\sigma_+$ and $\sigma_-$ polarized light, $\tilde{n}_{\pm}(\omega)$. The linearly polarized pump, which is a combination of $\sigma_{+}$ and $\sigma_{-}$ polarization, couples to both valleys equally.  The Stark shift of both valleys is identical, and so the difference in $\tilde{n}_{\pm}(\omega)$ is zero. This also explains why Kerr rotation is not sensitive to the polarization-independent background effects observed in the TR-R measurement.

The TR-KR spectrum shown in Fig.~\ref{fig:WSe2Kerr}b has a signal-to-noise ratio (defined as the ratio between the peak signal magnitude and the average uncertainty of each point) of 59, which is 5 times greater than the signal-to-noise ratio for the TR-R spectrum measured under the same pump and probe conditions.  This improved sensitivity is due to the suppression of background effects and the common mode noise rejection of the balanced photodetectors in TR-KR (Appendix~\ref{sec:methods}). With the parameters and collection times used in this experiment, the rotation angle sensitivity is about 3~\si{\micro\radian}. While this limit is not fundamental and can be improved with signal averaging, the collection parameters here are kept the same as for TR-R, in which sensitivity is more limited due to the background signal.

The Kerr spectrum is fit with a reflection model similar to the one used in Ref.~\cite{ye2016optical} in which the WSe$_2$ dielectric function is modeled with Lorentz oscillators (Appendix~\ref{sec:KerrAnalysis}). The analysis accounts for the multi-interface thin film effects from the Si/SiO$_2$ substrate that cause the TR-KR spectrum to vary non-negligibly from the $\Delta\kappa$ spectrum of WSe$_2$. By comparing the measured TR-KR spectrum (Fig.~\ref{fig:WSe2Kerr}b) to the $\Delta \kappa$ spectrum extracted from this fit (Fig.~\ref{fig:WSe2Kerr}c), we see qualitatively that the TR-KR measurement is primarily sensitive to changes in the imaginary part of the refractive index. This is in contrast to the TR-R measurement, which is primarily sensitive to changes in the real part (see SM). The $\Delta\kappa$ spectrum extracted from both the TR-R and TR-KR fits are plotted in Fig.~\ref{fig:WSe2Kerr}c.  Both show the characteristic ``Lorentzian-derivative" spectral dependence~\cite{sie2016observation,sie2015,kim2014ultrafast} with larger weight given to the low-energy side of the feature. This asymmetry has been observed previously~\cite{sie2015,sie2017,ye2016optical}, and was accounted for in our model by allowing the exciton resonance to broaden while preserving the oscillator strength of the transition~\cite{zimmermann1988dynamical,chemla1989excitonic,ell1989influence,knox1989femtosecond,ye2016optical}.
This broadening could be caused by inhomogeneous broadening due to the finite bandwidth ($\sim$ 14~\si{meV} FWHM) of the pump pulse~\cite{chemla1989excitonic,schafer1988theory,balslev1989two}, or from exciton-exciton interactions~\cite{sim2013exciton} between excitons generated by two-photon absorption.
From the fit to the Kerr spectrum, we find $\Delta E = 72 \pm 5$~\si{\micro eV}. The $\Delta E$ extracted from TR-KR and TR-R agree reasonably well, with overlapping 95\% confidence intervals. While a significant portion of the uncertainty in $\Delta E$ extracted from TR-R is due to measurement error in the TR-R spectrum, this is not the case for $\Delta E$ extracted from TR-KR. The TR-KR confidence interval is limited by the uncertainty in the initial Lorentz oscillator model parameters, and an even more precise result is achieved if these are well-known (more details of the fitting and analysis are in SM). The agreement between the TR-R and TR-KR fits using a simple two parameter model suggests that the salient features of the Stark shift phenomenon are detected by both methods. The improved sensitivity demonstrates that Kerr rotation is an effective probe of the valley-selective optical Stark effect in WSe$_2$.

\section{Optical Stark effect in M\MakeLowercase{o}S$_2$ probed by Kerr rotation}

The improved sensitivity of Kerr rotation to the valley-selective Stark effect is attractive for studying this phenomenon more broadly in TMDCs. In contrast to WSe$_2$, the exciton resonances in MoS$_2$ have significantly broader linewidths and lower oscillator strengths~\cite{li2014measurement}. This lessens the magnitude of the $\Delta R/R$ signal induced by the optical Stark effect, which is generally proportional to the slope of the exciton absorption feature~\cite{kim2014ultrafast} (Fig.~\ref{fig:OpticalStarkCartoon}c). Consequently, $\Delta R/R$ is reduced in MoS$_2$ and the background artifacts in TR-R are comparatively more significant, which may account for the lack of reports of valley Stark shift in MoS$_2$ so far.  The increased sensitivity of the TR-KR measurement allows us to definitively observe and quantify the valley-selective optical Stark effect in MoS$_2$.

\begin{figure}[bt]
\centering
\includegraphics{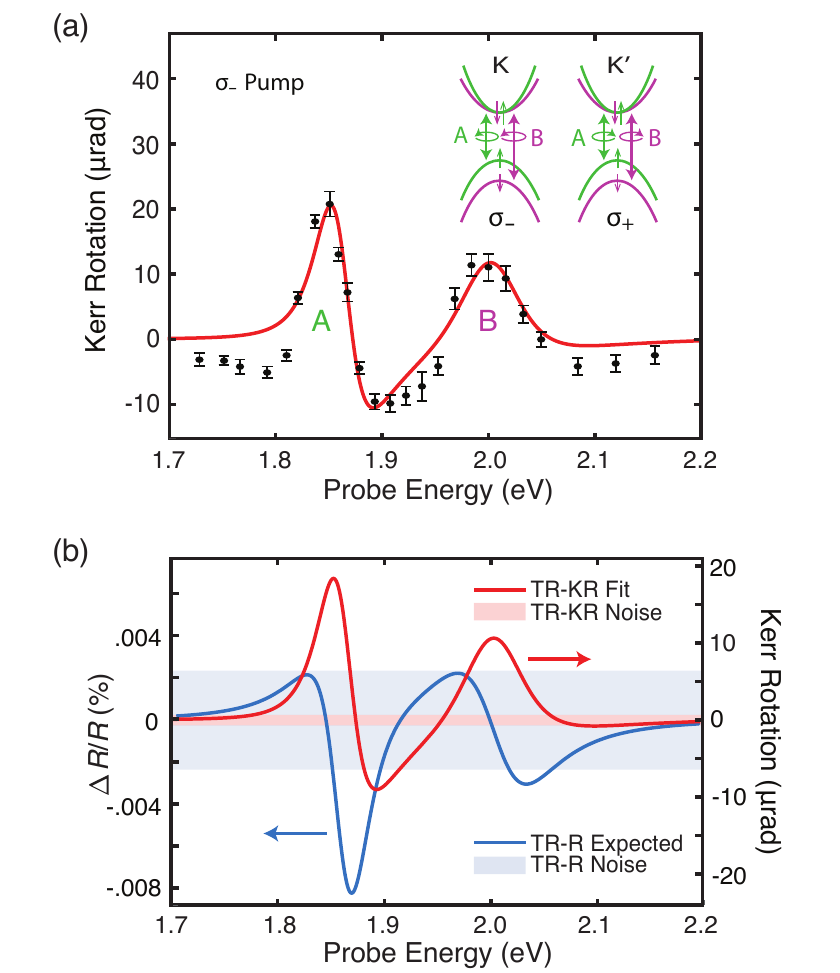}
\caption{Optical Stark effect in MoS$_2$ probed by TR-KR. (a) Kerr rotation angle as a function of probe energy in MoS$_2$ on Si/SiO$_2$ at 295~K using a $\sigma_-$ pump with $I_{\rm peak} = 320$~\si{MW/cm^2}, demonstrating optical Stark effect for both the A and B excitons. Inset: Schematic of the optical selection rules for MoS$_2$, including the spin-splitting that gives rise to the A and B excitons. (b) Fit to the measured Kerr spectrum (red) and the expected $\Delta R/R$ spectrum (blue) calculated from the $\Delta E$ extracted from the Kerr spectrum. The representative noise levels for the experiment are shown as shaded regions.}
\label{fig:MoS2Kerr}
\end{figure}

Fig.~\ref{fig:MoS2Kerr}a shows the TR-KR spectrum induced by the optical Stark effect in MoS$_2$, which exhibits a more complicated energy dependence compared to the straightforward WSe$_2$ TR-KR spectrum. This complexity originates from simultaneous significance of both the A and B excitons.  Previous work in WS$_2$ and WSe$_2$ reports the valley-selective Stark shift for only the A exciton~\cite{sie2017,sie2015,kim2014ultrafast,ye2016optical}. In these materials, the A and B excitons are spectrally distinct due to large spin-splitting in the valence band ($\sim 450$~meV)~\cite{zhu2011giant,cheng2013spin}; the spectral response near the lower-energy A exciton can be analyzed without considering the B exciton.  In MoS$_2$, however, the spin splitting is much smaller ($\sim 150$~meV), and both excitons must be included in the analysis. Because of the A and B exciton selection rules (Fig.~\ref{fig:MoS2Kerr}a inset), the $\sigma_-$ below-bandgap pump induces a shift in both the A and B excitons in the same K valley. While the optical Stark shift of two energetically distinct exciton states has been observed in few-layer ReS$_2$~\cite{sim2016selectively}, those excitons are not valley-sensitive~\cite{zhong2015quasiparticle}. The simultaneous shift of the A and B excitons in MoS$_2$ represents the first report of the energy- and valley-selective optical Stark effect in a single material system.

Demonstrated in Fig.~\ref{fig:MoS2Kerr}a, the MoS$_2$ Kerr spectrum can be understood using the same methods as in WSe$_2$ extended to include the blue-shift of both the A and the B excitons (Appendix~\ref{sec:KerrAnalysis}). Fig.~\ref{fig:MoS2Kerr}b shows the expected $\Delta R/R$ spectrum that corresponds with the fit to the Kerr spectrum. The expected $\Delta R/R$ signal for MoS$_2$ is comparable to the noise, whereas the Kerr spectral features are significantly larger than the noise floor. While additional signal averaging could reveal the TR-R spectrum, the TR-KR measurement for the same pump and probe conditions is a significantly more effective measurement. For MoS$_2$ at 295~K, the pump is red-shifted from the A and B excitons by $\Delta_{\rm pump}^\textrm{A} =$ 0.36~eV and $\Delta_{\rm pump}^\textrm{B} =$ 0.51~eV, respectively. We extract a Stark shift of $\Delta E_\textrm{A} = 8.4\pm1.3$~\si{\micro eV} for the A exciton and $\Delta E_\textrm{B} = 4.3\pm2.4$~\si{\micro eV} for the B exciton. If we assume that the proportionality constant $C$ is similar for the A and B excitons, then the inverse scaling with $\Delta_{\rm pump}$ predicts that $\Delta E_\textrm{B} \Delta_{\rm pump}^\textrm{B} = \Delta E_\textrm{A} \Delta_{\rm pump}^\textrm{A}$. We find that $\Delta E_\textrm{A} \Delta_{\rm pump}^\textrm{A} = 3.0 \pm 0.5$~\si{\micro eV\cdot eV} and $\Delta E_\textrm{B} \Delta_{\rm pump}^\textrm{B} =  2.2 \pm 1.2$~\si{\micro eV \cdot eV} agree within uncertainty, supporting this assumption.

This new observation in MoS$_2$ is compared to other TMDC monolayers in Table~\ref{tab:StarkConstants} by considering the proportionality constant $C$ extracted from available measurements of the A exciton valley-selective optical Stark shift in TMDCs.  While a direct comparison is not possible due to limited information on the pulse characterization, general trends are evident. There are large variations in $C$ (up to $10^3$) depending on both the material and the substrate, which is unsurprising since excitons in TMDC monolayers are very sensitive to dielectric environment~\cite{stier2016probing,buscema2014effect,latini2015excitons}.  Table~\ref{tab:StarkConstants} also lists the oscillator strengths ($f \propto \vert \mathcal{M} \vert^2$) extracted by fitting the reflectance contrast or absorption spectra. Since $C$ depends on exciton formation time and pulse width, which vary between materials and experimental setups, we do not necessarily expect $C$ to correspond numerically to $f$. However, when holding substrate constant, some correspondence does appear. For sapphire substrates in Table~\ref{tab:StarkConstants}, we see that $f_{\text{WSe}_2}  < f_{\text{WS}_2}$, and correspondingly $C_{\text{WSe}_2}  < C_{\text{WS}_2}$. Similarly, for Si/SiO$_2$ substrates $f_{\text{MoS}_2}  < f_{\text{WSe}_2}$ and $C_{\text{MoS}_2}  < C_{\text{WSe}_2}$. This trend $C_ {\textrm{MoS}_2} < C_ {\textrm {WSe}_2} < C_ {\textrm{WS}_2}$ also agrees with previous measurements of $f$ for the A exciton transition in these materials on fused silica ($f_{\text{MoS}_2}  = 0.301 < f_{\text{WSe}_2}  = 0.312 < f_{\text{WS}_2} = 0.554$)~\cite{li2014measurement}.

\begin{table}
\begin{ruledtabular}
\begin{tabular}{c c c c c }
Material & Substrate & Ref. & $f$ & $C$ (\si{eV^2 cm^2 / GW}) \\
 \hline
 WS$_2$ & Sapphire & ~\cite{sie2015} & 2.22 & $3.7 \times 10^{-3}$ \\
 WSe$_2$ & Sapphire & ~\cite{kim2014ultrafast} & 0.35& $3.6 \times 10^{-4} $  \\
 WSe$_2$ & Si/SiO$_2$ & ~\cite{ye2016optical} & 0.71 & $3.2 \times 10^{-5}$ \\
 WSe$_2$ & Si/SiO$_2$ & & 0.63 & $5.3 \times 10^{-5}$\\
 MoS$_2$ & Si/SiO$_2$  &  & 0.29 & $9.4 \times 10^{-6}$ \\
\end{tabular}
\end{ruledtabular}
\caption{Proportionality constant $C$ for optical Stark shift of A exciton in TMDCs estimated from literature. Our values, presented in the bottom two rows, calculate $C$ from the fits to Kerr rotation spectra. The magnitude of $C$ follows the same trend as the A exciton oscillator strength $f$ when holding substrate constant.}
\label{tab:StarkConstants}
\end{table}

\section{Conclusions}
\label{sec:conclusions}
This work demonstrates the use of Kerr rotation to measure the valley-selective optical Stark effect in monolayer TMDCs. By probing a different component of the complex refractive index, Kerr rotation dramatically improves the sensitivity to valley Stark shifts compared to reflectance-based techniques. The enhanced sensitivity of Kerr rotation allows detection of smaller Stark shifts with higher precision, enabling exploration of valley-selective optical Stark effects in other monolayer TMDCs and heterostructures. To this end, we exploit the improved precision to measure the previously unobserved optical Stark effect in MoS$_2$. The Stark shift of both the A and B excitons in MoS$_2$ are separable, demonstrating that energy- and valley-selective manipulation of excitonic states can be achieved simultaneously in a single material system. The measured Stark shift of $4$~\si{\micro eV} represents the smallest reported exciton energy shift in a TMDC. These results establish Kerr rotation as an accurate, high-precision probe of spin-valley degeneracy breaking and an attractive tool for exploring valley-selective differential energy shifts in monolayer and heterostructure devices.

\begin{acknowledgments}

This work was primarily supported by the Office of Naval Research under grant number N00014-16-1-3055 (valley manipulation). Sample preparation, characterization, and spectroscopy were supported by the National Science Foundation’s MRSEC program (DMR-1720139) at the Materials Research Center of Northwestern University. Chemical vapor deposition of monolayer MoS$_2$ was supported by the National Institute of Standards and Technology (NIST CHiMaD 70NANB14H012).
This work utilized Northwestern University Micro/Nano Fabrication Facility (NUFAB), which is partially supported by Soft and Hybrid Nanotechnology Experimental (SHyNE) Resource (NSF ECCS-1542205), the Materials Research Science and Engineering Center (NSF DMR-1720139), the State of Illinois, and Northwestern University.
This material is based upon work supported by the National Science Foundation Graduate Research Fellowship. H.B. acknowledges support from the NSERC Postgraduate Scholarship-Doctoral Program. N.P.S. gratefully acknowledges support as an Alfred P. Sloan Research Fellow.

\end{acknowledgments}
\appendix
\section{Experimental methods}
\label{sec:methods}
\subsection{Sample Preparation}
WSe$_2$ flakes are prepared from a commercial crystal by mechanical exfoliation onto a Si substrate topped with 285~nm of SiO$_2$. Samples are then annealed at 400$^{\circ}$~C for 3 hours in an Ar-rich environment. Monolayers are confirmed by PL measurements. All monolayer flakes have dimensions exceeding 30$\cross$80~\si{\micro\meter\squared}. Monolayer MoS$_2$ samples are grown by chemical vapor deposition directly on a Si/SiO$_2$ substrate~\cite{shastry2016mutual} (see SM). Ti/Au alignment windows are added to all samples using optical lithography and thermal gold evaporation to facilitate alignment to the area of interest.

\subsection{Time-Resolved Measurements}
All time-resolved measurements are made using a Ti:sapphire laser (Coherent Mira 900-F) with a repetition rate of 76~MHz and a nominal pulse width of $\sim$150~fs. The beam is split into a pump and a probe arm. For the pump, the 1.49~eV output from the Ti:sapphire laser is passed through a delay line and is chopped at 100~kHz using two crossed linear polarizers and a photoelastic modulator set to half-wave retardance. The probe arm  pumps an optical parametric oscillator (Mira-OPO) which provides tunable output from 1.65 -- 2.48~eV. The probe is mechanically chopped at 1.3~kHz. For the TR-R measurements, both the pump and probe beams are circularly polarized using a linear polarizer followed by a quarter-wave plate. The pump and probe beams are focused onto the sample using a $20\times$ objective to achieve overlapping spot sizes of 15~\si{\micro \meter} and 10~\si{\micro \meter}, respectively. The reflected probe is sent to an amplified Si photodetector (Thorlabs PDA10A) for lock-in detection using a time constant of 200~ms. Both the pump induced $\Delta R$ and the equilibrium reflectance $R$ are simultaneously measured for each probe energy. For the TR-KR measurements, the pump is circularly polarized while the probe is linearly polarized. The beams are again combined and focused onto the sample using a $20\times$ objective. The reflected probe is sent through a Glan-Thompson polarizing beam splitter, and the outputs are focused onto a balanced photodiode bridge (Thorlabs PDB450A) for lock-in detection. A half-wave plate just before the polarizing beam splitter is used to balance the photodiode signals. The photodiodes used for the TR-R and TR-KR measurements have comparable responsivity and transimpedance gain.  The noise floor in the TR-R and TR-KR measurements are $2\times 10^{-3}$~\si{\percent \per \sqrt \Hz} and $0.63$~\si{\micro \radian\per\sqrt\Hz}, respectively.

\section{Fit to Kerr rotation spectrum}
\label{sec:KerrAnalysis}
The complex Kerr rotation angle can be written as
\begin{equation}
\tilde{\theta} = \theta + i \xi
\label{eq:complexkerr}
\end{equation}
where $\theta$ is the Kerr rotation angle and $\xi$ is the Kerr ellipticity.  Assuming incident light is linearly polarized along the $x$-axis, and assuming that $\theta$ and $\xi$ are small, the Kerr angle can be written in terms of the reflection coefficients~\cite{kuch2015magnetic}:
\begin{equation}
\tilde{\theta} \approx \frac{r_{y}}{r_{x}} \implies \theta = \Re ( \tilde{\theta} ) = \Re \left(\frac{r_{y}}{r_{x}}\right)
\label{eq:complexkerrReflect}
\end{equation}
The reflection coefficients can be expressed in the circularly polarized basis:
\begin{equation}
r_x = \frac{1}{\sqrt{2}} (r_- + r_+), \quad r_y = \frac{-i}{\sqrt{2}} (r_- - r_+)
\label{eq:reflectioncoeffs}
\end{equation}
and plugging \eqref{eq:reflectioncoeffs} into \eqref{eq:complexkerrReflect} we have:
\begin{equation}
\theta = \Re \left( -i \frac{r_- - r_+}{r_- + r_+} \right) =  \Im \left( \frac{r_- - r_+}{r_- + r_+} \right)
\label{eq:KerrFitting}
\end{equation}

For a sub-resonant pump with $\sigma_-$ circular polarization, only the K valley will shift from the valley-selective Stark effect. Hence, the $\sigma_-$ component of the linearly polarized probe will probe the Stark-shifted K exciton, while the $\sigma_+$ component will probe the unperturbed K$^\prime$ exciton. At normal incidence, the polarized reflection coefficients $r_+$ and $r_-$ can be found from the Fresnel equations:
\begin{eqnarray}
\nonumber
r_{+} &= \frac{r_1 e^{-i(\beta_1+\beta_2)} + r_2 e^{i (\beta_1-\beta_2)} + r_3 e^{i(\beta_1+\beta_2)} + r_1r_2r_3 e^{-i (\beta_1-\beta_2)}}  {e^{-i(\beta_1+\beta_2)} +  r_1 r_2 e^{i (\beta_1-\beta_2)} + r_1 r_3 e^{i(\beta_1+\beta_2)} +  r_2r_3 e^{-i (\beta_1-\beta_2)}} \\
\nonumber
r_{-} &= \frac{r_1' e^{-i(\beta_1'+\beta_2)} + r_2' e^{i (\beta_1'-\beta_2)} + r_3 e^{i(\beta_1'+\beta_2)} + r_1' r_2'r_3 e^{-i (\beta_1'-\beta_2)}}  {e^{-i(\beta_1'+\beta_2)} +  r_1'  r_2' e^{i (\beta_1'-\beta_2)} + r_1' r_3 e^{i(\beta_1'+\beta_2)} +  r_2'r_3 e^{-i (\beta_1'-\beta_2)}}
\end{eqnarray}
with\[
r_{1} = \frac{\tilde{n}_{\textrm{air}}-\tilde{n}_{\textrm{WSe}_{2}}}{\tilde{n}_{\textrm{air}}+\tilde{n}_{\textrm{WSe}_{2}}} ,  r_{2} = \frac{\tilde{n}_{\textrm{WSe}_{2}}-\tilde{n}_{\textrm{SiO}_{2}}}{\tilde{n}_{\textrm{WSe}_{2}}+\tilde{n}_{\textrm{SiO}_{2}}} , r_3 = \frac{\tilde{n}_{\textrm{SiO}_{2}}-\tilde{n}_{\textrm{Si}}}{\tilde{n}_{\textrm{SiO}_{2}}+\tilde{n}_{\textrm{Si}}}
\] and \[
\beta_1 = 2 \pi \frac{\tilde{n}_{\textrm{WSe}_2} d_{\textrm{WSe}_2}}{\lambda}, \qquad \beta_2 =  2 \pi \frac{\tilde{n}_{\textrm{SiO}_2} d_{\textrm{SiO}_2}}{\lambda}
\]
where $\lambda$ is the wavelength of the light in vacuum, $d_{\textrm{WSe}_2} = 0.65$~nm and $d_{\textrm{SiO}_2} = 285$~nm are the material thicknesses, and $\tilde{n}_j = n_j + i \kappa_j$ is the complex index of refraction for material $j$. The primed $r^\prime_1$, $r_2^\prime$, and $\beta_1^\prime$ are found by replacing the index $\tilde{n}_{\textrm{WSe}_2}$ by the pump-modified $\tilde{n}_{\textrm{WSe}_2}^\prime$ index. We use literature values for the refractive index of Si and SiO$_2$ \cite{malitson1965interspecimen,vuye1993temperature} and the 500 \si{\micro m} Si substrate is approximated as semi-infinite.

The refractive index of WSe$_2$ in the region of the A exciton is modeled by a single Lorentz oscillator at the exciton resonance plus four additional oscillators centered at higher energies to account for off-resonance contributions. We note that this simple model does not account for the influence of lower-energy trions, which are valley-sensitive~\cite{wang2014valley,jones2013optical} and could therefore also exhibit a stark shift. Initial parameters of the Lorentz oscillators are found from a fit to the unperturbed reflectance contrast spectrum (see SM). To model the Stark shift in $\tilde{n}_{\textrm{WSe}_2}^\prime$, the energy and width of the oscillator associated with the A exciton is allowed to vary while preserving the oscillator strength~\cite{zimmermann1988dynamical,chemla1989excitonic,ell1989influence,knox1989femtosecond,ye2016optical}.  The finite width of the probe is accounted for by convoluting a Gaussian (FWHM$= 10$~\si{meV}) with the Kerr angle predicted by~\eqref{eq:complexkerrReflect} when fitting the measured data. This Gaussian convolution has only minor impact on the extracted $\Delta E$. For the MoS$_2$ analysis, $d_{\text{SiO}_2} = 325$~\si{nm}, and an additional oscillator is used to fit the B exciton resonance. The MoS$_2$ Kerr spectrum is fit by allowing the energy and width of \textit{both} the A and B exciton oscillators to vary independently, again while preserving oscillator strength.

\bibliography{LaMountain2017}

\end{document}